\documentclass[aps, pra, twocolumn, superscriptaddress, 10pt]{revtex4-2}

\usepackage[utf8]{inputenc}     
\usepackage{amsmath, amssymb}   
\usepackage{physics}            
\usepackage{siunitx}            
\usepackage{xspace}             
\usepackage{xcolor}             
\usepackage{soul}               
\usepackage{comment}            

\usepackage{multirow}   
\usepackage{booktabs}   
\usepackage{longtable}  
\usepackage{array}      
\usepackage{tabularx}   

\usepackage{graphicx}   
\usepackage{epstopdf}   

\setcitestyle{super} 
\usepackage[colorlinks = true,
            linkcolor  = blue,
            urlcolor   = blue,
            citecolor  = blue,
            anchorcolor= blue]{hyperref} 
            
\usepackage{etoolbox}
\patchcmd{\section}{\centering}{\raggedright}{}{}
\patchcmd{\subsection}{\centering}{\raggedright}{}{}

\usepackage[compact]{titlesec}
\titlespacing*{\subsection}{0pt}{0\baselineskip}{0\baselineskip}

\AtBeginDocument{
  \setlength\abovedisplayskip{0pt}
  \setlength\belowdisplayskip{0pt}
}

\AtBeginDocument{\RenewCommandCopy\qty\SI} 
\ExplSyntaxOn
\msg_redirect_name:nnn { siunitx } { physics-pkg } { none }
\ExplSyntaxOff

\begin{document}

\title{A scalable kinetic Monte Carlo platform enabling comprehensive simulations of charge transport dynamics in polymer-based memristive systems}

\author{Gerliz M. Gutiérrez-Finol}
\affiliation{Instituto de Ciencia Molecular (ICMol), Universitat de Val\`encia, Paterna, Spain}

\author{Kirill Zinovjev}
\affiliation{Departamento de Química Física, Facultad de Química, Universitat de Val\`encia, Burjassot, Spain}

\author{Alejandro Gaita-Ariño}
\email{alejandro.gaita@uv.es}
\affiliation{Instituto de Ciencia Molecular (ICMol), Universitat de Val\`encia, Paterna, Spain}

\author{Salvador Cardona-Serra}
\email{salvador.cardona@uv.es}
\affiliation{Departamento de Química Física, Facultad de Química, Universitat de Val\`encia, Burjassot, Spain}

\date{\today}


\begin{abstract} {Polymer-assisted ion transport underpins both energy storage technologies and emerging neuromorphic computing devices. Efficient modeling of ion migration is essential for understanding the performance of batteries and memristors, but it remains challenging because of the interplay of drift, diffusion, and electrostatic interactions, as well as the limitations of continuum and molecular dynamics approaches. Addressing these challenges is particularly relevant in the context of the climate and energy crisis, where high-performance, low-carbon technologies require optimized ion-conducting materials and devices. Here, we introduce a scalable and flexible stochastic simulation platform that uses Markov chain Monte Carlo methodology to model ion migration in solid-state systems. The platform employs a vectorized, rail-based representation of device geometry, enabling rapid simulation of lateral ion transport and space-charge effects while preserving the stochastic nature of hopping events. It accommodates a wide range of material systems and can integrate experimental input parameters without code modification. We also provide an implementation of the model that takes advantage of highly energy-efficient GPUs, improving the performance and reducing the carbon footprint of the simulations. Validation using polymer-based memristive devices demonstrates the simulator’s ability to capture key behaviors, including relaxation decay, current–voltage hysteresis, spike-timing-dependent plasticity, and learning/forgetting rates. By balancing computational efficiency with mesoscale physical considerations, the platform provides a versatile tool for exploring ion-driven phenomena in energy storage and neuromorphic devices, supporting exploratory research.}
\end{abstract}

\maketitle

\section*{1. Introduction: the Need for Scalable, General-Purpose Ion Transport Simulation}

Currently, scientists are facing interconnected global challenges, including the climate and energy crisis, the electrification of society, and the reduction of CO$_2$ emissions. At the same time, the rapid expansion of artificial intelligence (AI) and big data has led to algorithmic models of unprecedented complexity. Specifically focusing on this latter challenge, conventional von Neumann architectures suffer from intrinsic bottlenecks—in energy consumption, heat dissipation, and data transfer—that limit their scalability for large-scale, parallel workloads. \cite{Potla2022, Dhar2021} Although alternative non-von Neumann approaches have been explored, they remain insufficient for current computing requirements, especially since we need to simultaneously reduce our collective energy consumption. This motivates the search for new paradigms such as in-memory and neuromorphic computing. \cite{Zhu2023}

Neuromorphic computing seeks to reproduce the energy-efficient information processing of the brain by employing synthetic materials to mimic neuron-like behavior. In recent years, proposals from chemistry, physics, electronics, and materials science have advanced the design of artificial neurons and neural networks. 

From the perspective of materials design, ion migration plays a central role in both neuromorphic computing and energy technologies. In batteries and supercapacitors, this phenomenon governs charging rates, cycle life, and processing performance. \cite{Sand2025, Radjendirane2024} In information technologies, ion migration is one of the most successful approaches to obtain memristive behavior and neuromorphic devices, where controlled ion motion enables non-volatile memory and synaptic plasticity. \cite{Kwak2024, Park2025, Ji2025} Unlike the relatively free motion of ions in liquids, solid-state transport proceeds via hopping between vacancies, interstitials, or disordered pathways in crystalline or amorphous matrices. Understanding these mechanisms is therefore essential to improving both energy storage and neuromorphic devices. \cite{Chen2025, Yang2022, Huang2025}

Ion migration in solid-state materials arises from the combined action of drift, diffusion, and electrostatic interactions. Under an applied bias, drift dominates: mobile ions move in response to the electric field, with cations driven toward the negative electrode and anions toward the positive. As ions accumulate at interfaces, Coulombic repulsion between like charges builds up, producing an internal electric field that opposes the applied bias. This effect can reduce the net driving force for drift, promote space-charge formation, and trigger a transition toward diffusion-controlled transport.

When the bias is removed, transport is governed mainly by diffusion as ions relax from the high-density interfacial regions toward a more uniform distribution. However, Coulombic effects persist: densely packed ions still resist further compaction, slowing the return to equilibrium. \cite{Alvarez2024} These dynamics—fast drift-driven accumulation followed by slower, repulsion-modulated relaxation—are observed in memristive devices, perovskite films, and other ionic conductors. \cite{Chen2025, Yang2022, Huang2025} They can significantly impact macroscopic observables such as hysteresis in current–voltage curves, charge retention times, and overall transport efficiency. \cite{Kwantwi2024, Wolanska2024}

However, simulating ion transport remains challenging: continuum approaches, such as drift–diffusion simulation, cannot be used as a routine method because of the vastness of model parameters. \cite{Diethelm2025} Furthermore, some relevant ion-related processes are often not well understood and are typically not introduced into these models, specifically regarding interactions with transport layers and ion penetration effects. \cite{Diethelm2025} Molecular dynamics (MD) captures atomic detail but becomes prohibitive for large systems or long timescales, typically up to a few nanoseconds (ns), particularly when rare events such as long-range hopping dominate transport, potentially spanning microseconds ($\mu$s). \cite{Deng2023}

Alternatively, modeling tools like Molecular Dynamics (MD) are employed to study ion movement in nanoscale fluid systems, such as investigating ion drift in nanochannel water flows under external electric fields for desalination. \cite{Sofos2020} Conversely, for simulating kinetic phenomena in solid-state materials, advanced software solutions like kMCpy, a python package for kinetic Monte Carlo (kMC), are developed to compute ionic transport properties in crystalline materials. KMC methods, unlike MD, are better suited to access longer timescales (up to $\mu$s or ms) and larger length scales (up to $\mu$m) in these crystalline systems. \cite{Deng2023}

The simulator presented here is fully vectorized for efficiency and introduces a \emph{rail}-based abstraction of device geometry: ions migrate stochastically along predefined one-dimensional paths (“rails”), each carrying a fixed number of mobile ions. Vertical or inter-rail motion is excluded, providing a controlled dimensionality reduction that preserves device-level transport statistics while enabling scalable simulations. Increasing the number of rails improves sampling without altering the stochastic rules. Experimental data can be directly imported via external files, and the simulator is available in both \texttt{MATLAB} and \texttt{Python}. With flexible rate modeling and a modular design, it accommodates diverse materials and transport mechanisms. Validated through memristive device case studies, it qualitatively reproduces experimental behaviors and provides microscopic insight into resistive switching and ion migration. By balancing efficiency, scalability, and reproducibility, this tool supports research on ion-driven phenomena across energy storage, non-volatile memory, and neuromorphic computing. It is part of an effort towards frugal modelling in molecular electronics.\cite{Gutierrez2025}

\section*{2. Previous Theoretical Background}

At the microscopic scale, as mentioned in the introduction, ionic transport in solids arises from several mechanisms that frequently coexist and influence each other. Under an applied electric field, drift dominates, causing ions to move directionally and accumulate at interfaces, which can significantly affect device performance and stability \cite{Garcia2022, Kemp2021}. This drift is particularly susceptible to field-enhanced ion mobility at the enormous field strengths found in nanoscale devices. \cite{Kemp2021} Simultaneously, hopping, or jump diffusion, allows ions to thermally activate and move between discrete lattice sites via vacancies or interstitials; this process is sensitive to temperature, lattice symmetry, and local chemical environment, and may display correlated behavior in disordered regions \cite{Singh2025, Eames2015}. The local energy landscape around the migrating ion determines the ease of migration. \cite{Singh2025} In addition, trapping and detrapping can temporarily immobilize ions at deep energetic sites such as defects or grain boundaries, \cite{Singh2025, Zuo2022} followed by release through thermal or electrical activation, adding further temporal complexity. Finally, in the absence of external fields, diffusion becomes the dominant mechanism, driving ions from regions of high to low concentration according to Fick’s laws \cite{Yantara2024, Zuo2022}. Together, these mechanisms interact to determine effective ionic conductivity and transport anisotropy, emphasizing the need for simulation approaches that can accurately capture their interplay within computationally feasible models.

Modeling ionic motion in solid-state and membrane systems has traditionally relied on coupled differential equations, most notably the drift--diffusion or Nernst--Planck formulations, sometimes augmented with Poisson's equation to account for electrostatic effects. While these continuum approaches are computationally efficient, solving the equations often encounters issues of numerical stability and convergence, and they cannot easily capture correlated or collective migration events that govern device performance \cite{Canepa2025, Deng2023}, relying instead on macroscopic parameters that neglect the stochastic nature of ionic random walks \cite{Canepa2025}. Furthermore, these treatments struggle when the migration path is curved, such as the path anion vacancies follow in the perovskite structure \cite{Kemp2021}, or when ion transport involves non-linear mobility at high electric field strengths (e.g., MV cm$^{-1}$) \cite{Kemp2021}, which invalidates the linear assumptions of the models. This is because the trajectory of the migrating anion is curved, so the contribution of the applied electric field varies during the jump \cite{Kemp2021}.

\subsection*{Differences between charge transport and ion migration in modeling terms}

Considering the similarities between polymer-based memristive systems and the classical structure and composition of Light Emitting Electrochemical Cells (LECs), we can establish a parallel understanding of the models developed for the former. This allows extraction of electronic properties that yield neuromorphic behavior.

An important feature in the modeling of LECs is the distinction between electronic charge transport and ion migration. The interaction between these two phenomena has historically been challenging to describe, leading to the development of two principal models: the electrodynamic (ED) model and the electrochemical doping (ECD) model.

The ED model, introduced by de Mello and colleagues \cite{DeMello1998, DeMello2002}, attributes enhanced charge injection to the formation of electric double layers (EDLs) at the electrodes, which screen the active layer from the external field. Here, steady-state currents are dominated by electronic diffusion, with negligible drift; no internal p–n junction is predicted within the bulk material. Conversely, the ECD model, developed by Pei et al. \cite{Matyba2009, Smith1997, Pei1995}, proposes that injected charges trigger bulk doping of the conjugated polymer via ion migration. This results in expanding p- and n-type regions that eventually meet to form a dynamic p–n junction, where light emission occurs. Evidence supporting this model includes spatially resolved photoluminescence quenching, scanning Kelvin probe microscopy (SKPM) potential profiling, and transient conductivity measurements.

To bridge these perspectives, unified drift–diffusion–Poisson models \cite{Van_Reenen2010} have been introduced, simultaneously describing electrons, holes, and ions through coupled non-linear PDEs. While conceptually powerful, these models require careful numerical treatment to maintain stability and convergence, and their dimensionality grows rapidly when including realistic physical factors such as injection barriers, disorder, recombination, and morphology-dependent mobilities. This makes them computationally costly and less suited for rapid parameter sweeps or adaptation to new experimental scenarios.

\section*{3. KMC methodology for ion transport}

KMC has emerged as a powerful tool for bridging microscopic realism with computational efficiency in ion transport simulations. By reducing atomic dynamics to stochastic event-based hops—with rates derived from transition state theory or first-principles methods—KMC efficiently captures rare events and long-timescale processes that are inaccessible to molecular dynamics. Its formulation as a Markov chain, where transitions depend only on the current state, enables exploration of complex energy landscapes through memoryless random walks. This makes KMC particularly well-suited for studying resistive switching, filament formation, and diffusion-driven dynamics in realistic devices, extending simulation times far beyond those achievable with conventional atomistic techniques. \cite{Canepa2025, Kopperberg2025, Aldana2023, Deng2023}

This method simulation requires a discrete set of ion positions—energetically distinct sites in the host material—and the migration barriers separating them. These barriers, obtained from first-principles calculations (e.g., density functional theory) or methods such as the nudged elastic band technique, determine the ease of ion motion. The transition rate for each event is calculated using Arrhenius-type expressions:
\vspace{1em}
\begin{equation}\label{eq:hop_rate}
\Gamma = \nu^* \exp\left(-\frac{E_m}{k_{\text{B}}T}\right)
\end{equation}
\vspace{-1em}

Where $\nu^*$ is the attempt frequency, $E_m$ the migration barrier, $k_{\text{B}}$ Boltzmann’s constant, and $T$ the temperature. In crystalline materials, cluster expansion techniques can refine site energies and barriers by explicitly accounting for the local chemical environment. \cite{Canepa2025, Deng2023}  

At each KMC step, all possible migration events—such as hops, trapping, or detrapping—are listed with their rates. After summing the total rate, one event is randomly chosen in proportion to its probability, and the simulation time is incremented by a stochastic interval drawn from an exponential distribution. By repeating this process over millions of steps, KMC yields statistically robust transport properties such as diffusivity and conductivity, even under far-from-equilibrium conditions. \cite{Andersen2019, Deng2023} Unlike classical Monte Carlo, which samples equilibrium states, KMC directly captures time-dependent dynamics, making it particularly effective for ion migration studies on experimentally relevant timescales. \cite{Andersen2019, Saunders2020}

Despite these strengths, most existing KMC implementations are specialized and difficult to generalize, which limits reproducibility and broader applicability. They often lack modularity, making it challenging to adapt to different material classes such as amorphous systems, composites, or defect-rich structures. \cite{Deng2023} Furthermore, as materials science increasingly relies on high-throughput and multiscale modeling, there is a growing demand for frameworks that can seamlessly interface with first-principles calculations while remaining computationally efficient. Rising computational demands also emphasize the need for sustainable algorithms: efficient, low-overhead simulations not only reduce environmental impact but also make advanced modeling capabilities more widely accessible. 

To address these challenges, we introduce a scalable stochastic simulation platform for modeling ion migration in solid-state systems. Building on concepts from Kinetic Monte Carlo and Markov chain theory—and inspired by our earlier work on lanthanide molecular nanomagnets \cite{Gutierrez2023}—the framework employs a probabilistic, event-based methodology adapted to a fundamentally different class of materials. It integrates systematic event generation, efficient rate calculation, and modular event handling to capture stochastic ion dynamics across a broad range of systems. While it does not reproduce exact atomic trajectories, the approach balances physical realism with computational speed, avoids the overhead of coupled PDE solvers, and reveals clear qualitative trends. Validated here for memristive materials, the platform is general and extensible to other device architectures, ensuring adaptability and scalability for future applications. \cite{Gutierrez2023, Gutierrez2025, Canepa2025, Deng2023}

\subsection*{Mapping to the code}

To incorporate electrostatic interactions, ion migration is governed by site-specific, time-dependent probabilities derived from local effective fields. At each step, the field is computed as the sum of the applied bias and Coulombic contributions from nearby ions; in dense regions, the resulting repulsion lowers the net field and reduces drift into that area. From this effective field, the energetic cost of a hop is obtained, and the hopping probability is assigned through a Boltzmann factor,
\vspace{1em}
\begin{equation}\label{eq:Boltzmann_distribution}
P_{\text{hop}} \propto \exp\left(-\frac{\Delta E}{k_{\text{B}}T}\right)
\end{equation}
\vspace{-1em}

Forward and backward probabilities are normalized to preserve detailed balance in the absence of bias, allowing drift and diffusion to arise naturally from the same rules while space-charge effects follow from the dynamic modification of local fields. Implemented in a vectorized and parallelizable fashion, the simulator efficiently reproduces the interplay of drift, diffusion, and Coulombic interactions without explicitly solving Poisson’s equation (see Supplementary Information, section S2).

As indicated above, to benchmark the platform, we focused on memristive materials, particularly polymer-based organic devices where resistive switching involves a complex interplay of ionic and electronic processes. In these systems, ion migration couples with charge trapping, interfacial barrier modulation, filament formation, and redox reactions, all of which shape switching kinetics and stability. \cite{Prado2022, Zhang2025, Zhang2021, Hoch2025, Khan2024} Their multi-mechanism behavior makes them an ideal test case for validating the simulator against experimentally observed trends, ensuring that the model captures both qualitative patterns and quantitative transport characteristics.


\section*{4. Simulation Results}
The stochastic kinetic Monte Carlo framework developed in this work successfully reproduces the mesoscale transport behavior of polymer-based memristive devices under different operating regimes. In particular, to assess the versatility of the simulator, we modeled three representative experimental protocols: relaxation decay, current–voltage hysteresis, and potenciation/depression tests. \cite{Prado2022} In each case, the aim was not to replicate every microscopic detail of the device, but rather to capture the dominant mesoscale processes drift, diffusion, and space-charge effects that govern the devices performance's. Geometry, ionic density, and operating conditions were chosen to reflect typical parameters for polymer composite memristive devices reported in the literature, allowing us to obtain a meaningful comparison with existing experimental benchmarks while still preserving a frugal computational cost.

To further improve computational and energy efficiency of the platform and thus facilitate a frugal approach, we implemented an alternative version of the simulation code using the JAX\cite{jax2018github} framework. Modern GPUs, with their massively parallel architectures, are inherently more energy-efficient than CPUs for highly parallelizable numerical tasks. This efficiency arises from the high throughput per watt achieved by streaming multiple processing units optimized for vectorized operations, which is particularly beneficial for heavy linear algebra workloads. The JAX implementation closely follows the original NumPy-based code but leverages GPU acceleration. All array operations were replaced with their JAX equivalents, allowing execution on either CPU or GPU backends without major structural changes. To further improve runtime performance, we employed just-in-time (JIT) compilation, which compiles the computational kernels into optimized machine code, significantly reducing Python overhead during repeated evaluations. This approach provides substantial performance gains while being consistent with the reference NumPy implementation. The quantitative comparison between CPU and GPU performance and energy metrics is summarized in Tables S2-S3 of the Supplementary Information, showing that GPU execution consistently outperforms the CPU implementation both in runtime and energy efficiency, achieving up to a 17-fold speedup and over 7-fold energy savings for large-scale simulations.

\begin{figure}[h!]
    \centering
    \includegraphics[width=1.0\linewidth]{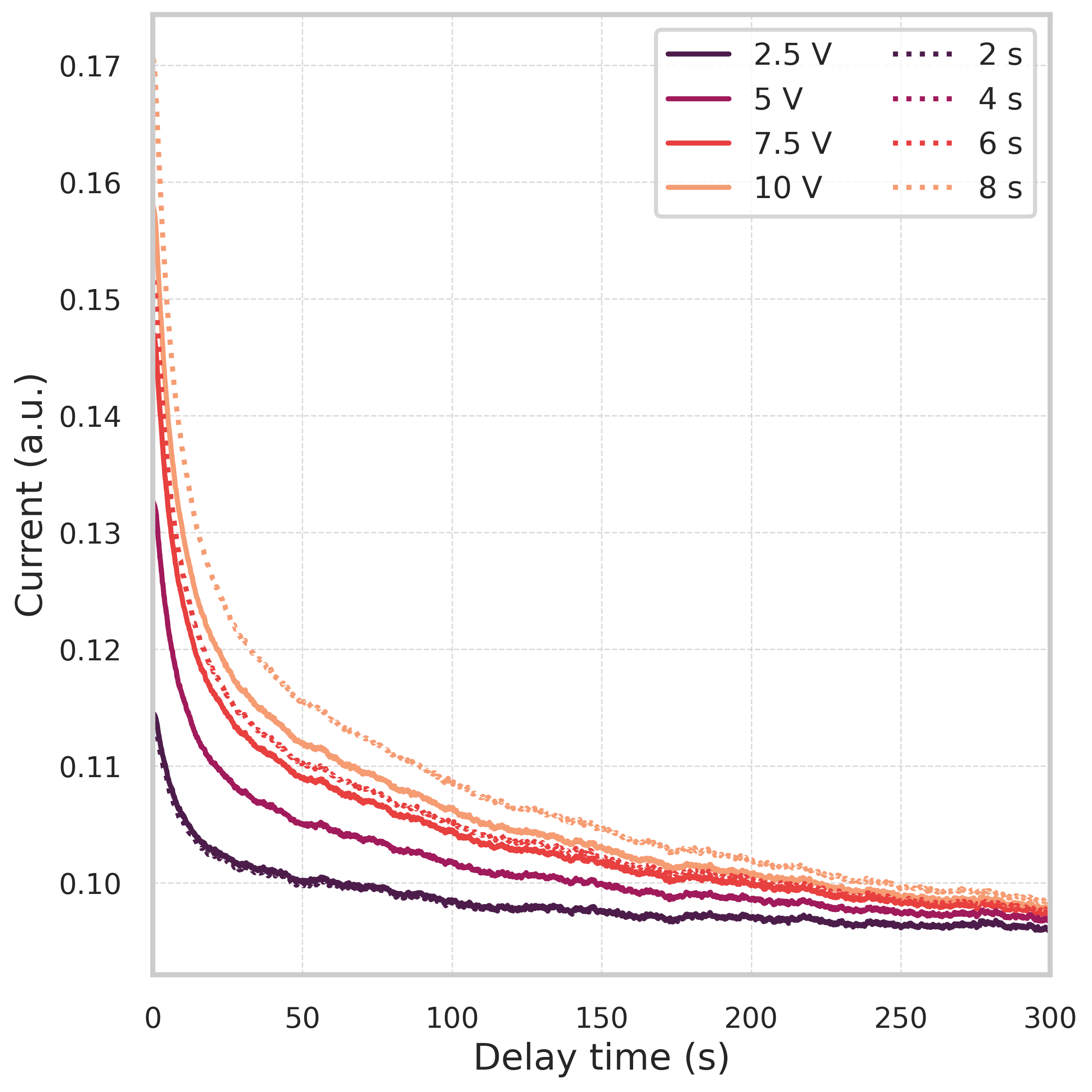}
    \caption{Simulated decay of current at different polarization times and voltages. The plotted signal (a.u.) corresponds to the ratio between the number of ions located in the active decile and the total number of ions, computed at each simulation step. Solid curves represent simulations with varying polarization voltages (2.5–10 V) at a fixed polarization time of 4 s, whereas dotted curves correspond to varying polarization times (2–8 s) at a fixed voltage of 5 V. Increasing the polarization voltage or time leads to a higher initial ratio and a slower relaxation process, indicating stronger ionic polarization effects.}
    \label{fig:relaxation}
\end{figure}

 In the relaxation decay protocol, the model starts from a rapid ion migration toward electrode interfaces under an applied bias, followed by a slower, repulsion-modulated redistribution once the bias was removed. Tracking individual ion positions over time showed the evolving spatial density gradients, which represents a direct view of how local electric fields get weak as space-charge regions build up. In Figure~\ref{fig:relaxation} the relaxation dynamics following the removal of the initial bias are presented.

The simulations begin from a uniform ionic distribution in which approximately one-third of the available sites are initially occupied, with ions randomly positioned across the device. During the polarization stage, a constant voltage is applied for a prescribed duration, and the curves in the figure reflect these two experimental controls: solid lines correspond to different applied voltages (2.5–10 V) at a fixed polarization time, while dotted lines correspond to different polarization durations (2–8 s) at a fixed voltage. Once this charging period ends, the voltage is removed and the system evolves freely, allowing the ionic distribution to relax toward equilibrium. Under these conditions, the fraction of ions accumulated near the electrode interface decreases exponentially with time, which indicates a single diffusion-controlled relaxation process once field drift is no longer dominant. Regardless of the applied voltage or polarization time, all curves converge to a similar asymptotic value, demonstrating the system’s return to a homogeneous distribution.
 
 It is important to note that, at short enough times, the system can present short-time memory (STM), corresponding to noticeable values of current at times $<10$~s and applied fields lower than 5~V. Increasing the voltage or the duration of the applied field can turn this effect into long-time memory (LTM), which persists up to 45~s. Regarding the relaxation dynamics, a single polarization phase of a few volts and a few seconds still corresponds to the linear response regime. This is especially true for the time, where we see a noticeable proportionality between the number of ions accumulated in the first decile and time. At moderate voltages ($<5$~V), the ionic population scales linearly with initial pulse duration, confirming operation within the quasi-linear response regime. Deviations appearing at higher voltages ($>7.5$~V) reveal the onset of field-induced saturation and Coulombic crowding, which limit further ionic drift and enhance the consequent Coulombic effects. In this regard, it can be suggested that, in the absence of sizeable repulsion between ions, we do not expect significant changes when moving from weaker voltages or shorter times. These results demonstrate that the model correctly captures the transition from drift-dominated to diffusion-limited transport, a key feature of ion-conducting polymers.

 Monitoring the spatial and temporal dynamics of ionic charge carriers is central to the hysteresis simulations, where the program captures how the history of ionic accumulation shapes the width and asymmetry of the I–V loop under different sweep speeds and bias polarities.
 
\begin{figure}[h!]
    \centering
    \includegraphics[width=1.0\linewidth]{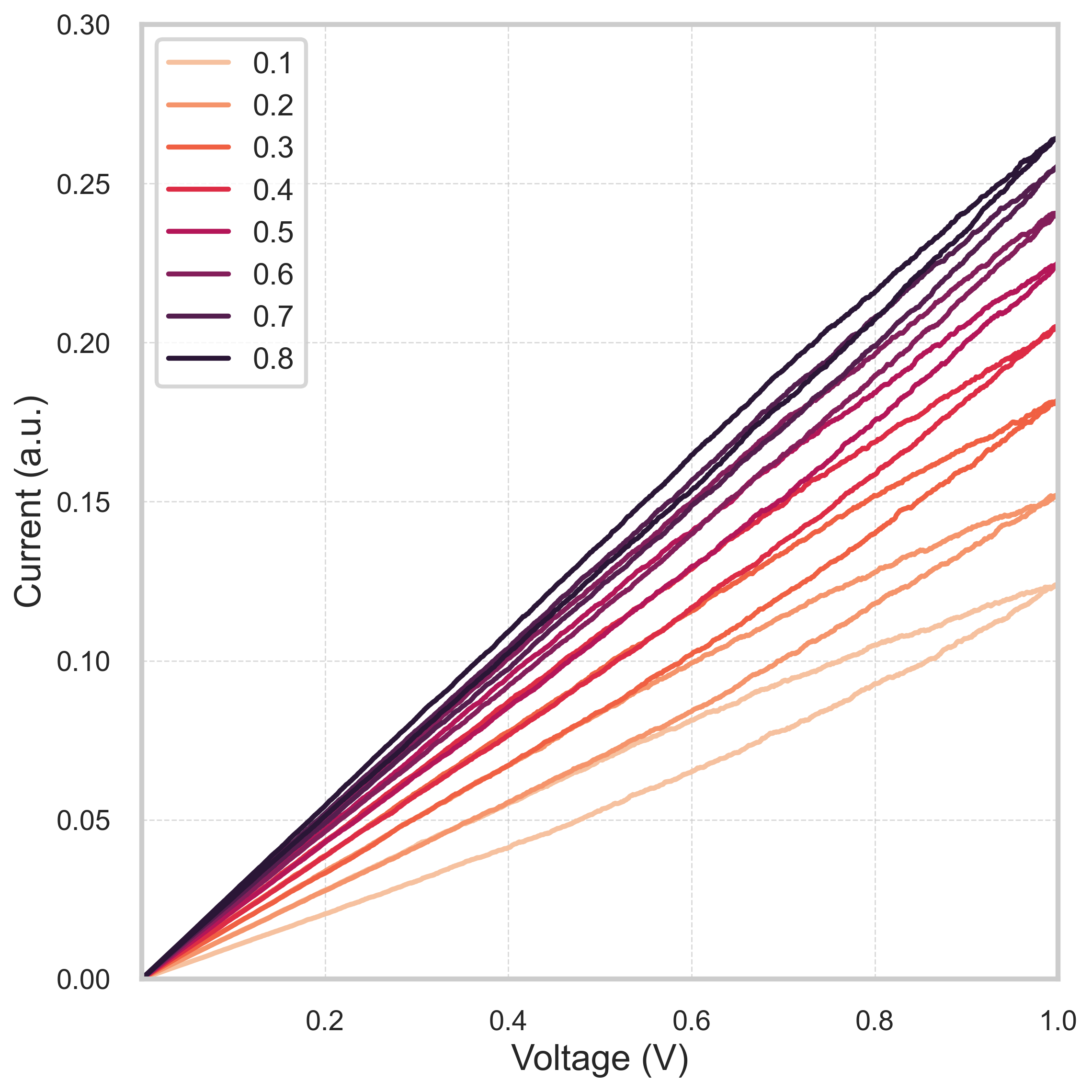}
    \caption{Simulated I--V hysteresis loops obtained at different voltage sweep rates ($0.1$--$0.8$ V/s). Each sweep begins from the final ionic configuration reached at the end of the preceding sweep, thereby capturing the cumulative influence of sweep history on the device response. The simulated current (a.u.) corresponds to the decile-based ionic population used in the model: during forward (positive-voltage) segments, the active population is taken as the first decile, whereas during reverse (negative-voltage) segments it is taken as the tenth decile, following the selection rule prescribed by the simulation algorithm.}

    \label{fig:hysteresis}
\end{figure}

 The current–voltage characteristics depicted in Figure~\ref{fig:hysteresis} further validate the model’s ability to capture the interplay between ionic motion and electrical response. The simulated I–V loops exhibit pronounced hysteresis, which is a well-known characteristic of memristive devices. In this case we also observe that the width and asymmetry present a strongly dependence  on the voltage sweep rate. At slow sweeps ($<$ 0.4~V\,s$^{-1}$), ions have sufficient time to migrate and accumulate at the interfaces, leading to strong polarization effects and broad hysteresis loops as expected. As the sweep rate increases, the ionic subsystem proves to be unable to follow the rapidly varying alternate electric field, resulting in progressively narrower and more symmetric loops. This rate-dependent effect is again a physical representation of the competition between electronic conduction and time-dependent ionic drift. Again, a demonstration of the power of the simulator to reproduce memory effects, with the voltage history dependence arising intrinsically from ion redistribution.

\begin{figure}[h!]
    \centering
    \includegraphics[width=0.90\linewidth]{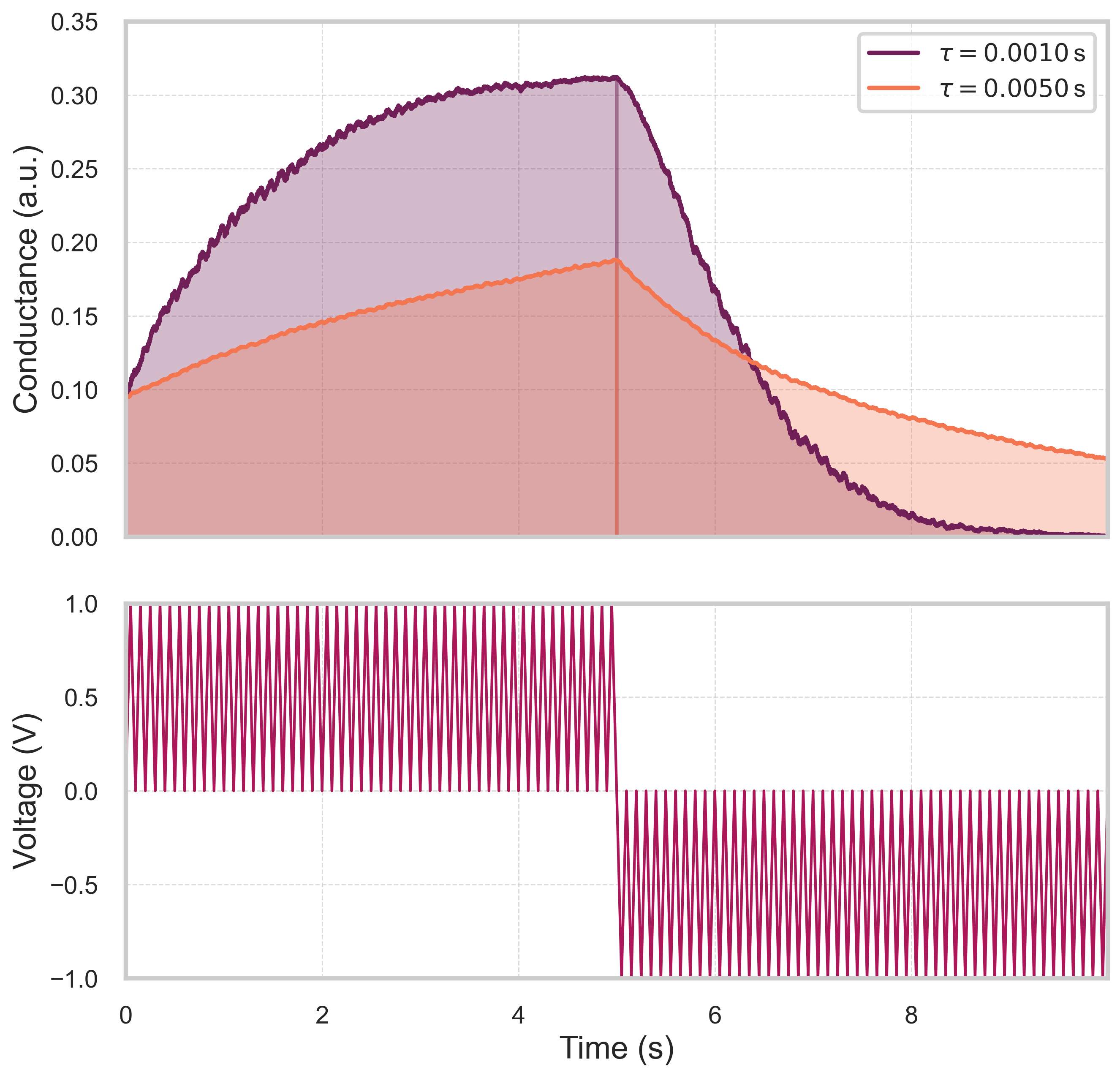}

    \caption{Simulated potentiation--depression responses for two ionic relaxation times ($\tau = 0.0010\,\text{s}$ and $\tau = 0.0050\,\text{s}$). The upper panel shows the evolution of device conductance during the learning (left) and forgetting (right) phases, with shorter relaxation times producing sharper potentiation and faster decay. The lower panel shows the applied voltage waveform used to drive the dynamics.}

    \label{fig:potdep}
\end{figure}

Beyond continuous voltage application, the simulator is also able to capture the fine dynamic conductance modulation to resemble synaptic plasticity in typical neuromorphic materials. Figure~\ref{fig:potdep} shows the potentiation–depression behavior obtained under alternative packages of voltage pulses for two representative relaxation times. The conductance increases during positive stimulation (potentiation phase) and decreases upon bias reversal (depression phase), forming an asymmetric cycle typical of a memristive-like learning process. The degree of asymmetry and memory retention specially depends on the relaxation constant: short relaxation times ($\tau < 1\times10^{-3}$~s) produce fast conductance changes with labile persistence of the memory effect, whereas increasing relaxation times ($\tau > 5\times10^{-3}$~s) result in slower, more persistent responses. This tunability highlights how ionic relaxation governs the balance between short- and long-term memory analogs.

\section*{Conclusion and context}

The simulator's predictions align with key qualitative trends from experimental reports on polymer-based memristive materials, including the sensitivity to different relaxation times, bias amplitude, and the sweep-dependence of the hysteresis loop shape. 
Because the implementation is computationally lightweight-scaling nearly linearly with the number of ions and time steps, large parameter sweeps can be performed in seconds to minutes on a standard workstation. This enables rapid and frugal exploration of how electric field strength, morphology, and materials properties jointly influence ionic behavior, offering a practical complement to more resource-intensive techniques, and perhaps avoiding them in some cases.

These results demonstrate that the kinetic Monte Carlo framework provides a physically grounded yet computationally efficient description of ionic migration in polymeric memristive materials. Using experimentally obtainable fitting parameters, the model reproduces essential observables such as relaxation decay, hysteresis modulation, and asymmetric potentiation–depression cycles using a unified stochastic mechanism. The agreement between these simulated trends and the bibliographic experimental behaviors confirms that the approach captures the correct hierarchy of time scales governing drift, diffusion, and electrostatic screening. This makes the simulator a powerful exploratory tool for correlating device architecture, operating conditions, and material properties with emergent resistive switching behavior, supporting rational design of next-generation neuromorphic components.

It is within this context that the model presented in this work, introduces a significant improvement. By adopting a probabilistic treatment of the relevant physical events, we have eliminated the neeed on globally convergent differential equation solvers as well as realistic experimental descriptors ca be used in the simulation. The approach presented here allows frugal modelling and adds extra flexibility for the inclusion of new physical mechanisms without destabilizing the model. One could for example include the charge-trapping effect of scarce defects by simply altering the probabilities governing a handful of random positions. Using the MC model, we are able to simulate the spatiotemporal evolution of polymeric ion migration devices and their operation under realistic conditions with a robustness and explanatory power not accessible to previous frameworks.

\label{conclusion}

\section*{Data Availability}

All datasets generated and analyzed in this study are openly provided in the accompanying repository. Raw and processed files required for full reproducibility are available in the ``data for reproducibility'' directory within 
\href{https://github.com/gerlizg/CupFlow}{github.com/gerlizg/CupFlow}.

\section*{Code Availability}

All simulation code used in this study is openly available in the CupFlow repository at 
\href{https://github.com/gerlizg/CupFlow}{github.com/gerlizg/CupFlow}. The repository contains the full kinetic Monte Carlo framework, example scripts, and all configuration files required to reproduce the results presented in this work.


\bibliographystyle{vancouver}
\bibliography{bibliography}

\section*{Acknowledgements}
A.G.A. has been supported 
by the Generalitat Valenciana (GVA) CIDEGENT/2021/018 grant. 
A.G.A. thanks grant PID2020-117177GB-I00 funded by MCIN/AEI/10.13039/501100011033 (co-financed by FEDER funds). 
This study is part of the Quantum Communication programme and was supported by grant PRTR-C17.I1 funded by MCIN/AEI/10.13039/501100011033 and European Union NextGenerationEU/PRTR, and by GVA (QMol COMCUANTICA/010).

\section*{Conflict of Interest}
The authors declare no competing interests.

\end{document}



\title{Supplementary information for ``A scalable kinetic Monte Carlo platform enabling comprehensive simulations of charge transport dynamics in polymer-based memristive systems''}

\author{Gerliz M. Gutiérrez-Finol}
\affiliation{Instituto de Ciencia Molecular (ICMol), Universitat de Val\`encia, Paterna, Spain}

\author{Kirill Zinovjev}
\affiliation{Departamento de Química Física, Facultad de Química, Universitat de Val\`encia, Burjassot, Spain}

\author{Alejandro Gaita-Ariño}
\email{alejandro.gaita@uv.es}
\affiliation{Instituto de Ciencia Molecular (ICMol), Universitat de Val\`encia, Paterna, Spain}

\author{Salvador Cardona-Serra}
\email{salvador.cardona@uv.es}
\affiliation{Departamento de Química Física, Facultad de Química, Universitat de Val\`encia, Burjassot, Spain}

\date{\today}

\maketitle

\tableofcontents

\newpage

\section{Device Geometry and Material Parameters}

In this section, we describe the structural and physical assumptions used to model the polymer-based memristive device. The aim is to bridge the gap between the experimental configuration and its computational representation, ensuring that the simulated system retains the essential spatial scales, material composition, and boundary conditions of the real device.

To reproduce the experimental behavior of a polymer-based memristive device in simulation, we began with a computationally simple but physically representative model. Rather than attempt to capture the full complexity of coupled ionic–electronic processes, we focused on the motion of cations within the active layer. The polymer matrix and anions are treated implicitly as a fixed background that defines the energy landscape and electrostatic environment for the mobile cations. The following subsections detail the simulated structure, the way it is discretized, and the boundary conditions applied during calculations.

\subsection{Simulated polymer-based memristor structure}

The simulated structure is based on the experimental device reported by Prado-Socorro et al. \cite{Prado2022} The active layer (experimental device) is a composite of Super Yellow (polyphenylenevinylene derivative, SY) as the semiconducting host, Hybrane DEO750 8500 as an ion-conducting polymer, and lithium triflate (\ce{LiCF3SO3}) as the ionic dopant. This combination forms a solid-state ionic conductor capable of exhibiting memristive behavior.

For the purposes of the present simulation, only lithium cations (\ce{Li+}) are considered as mobile species. Anions are assumed to be effectively immobile on the timescales of interest, serving as a fixed counter-charge background. This simplification reduces computational complexity while focusing on the dominant ionic process observed experimentally-\ce{Li+} migration and accumulation under bias.

\subsection{Lattice Dimensions and Discretization Scheme}

The active layer (simulated device) is represented as a two-dimensional lattice, with the vertical direction corresponding to discrete “rails” across the device thickness and the horizontal direction corresponding to sites available along each rail. Each lattice site represents a possible position for a single cation within a nanoscale region of the polymer matrix.

The total number of cations is set to match a chosen fraction of the available lattice sites, based on the experimental ion concentration. In practice, this concentration is understood as the ratio of ionic species to the host polymer matrix (i.e., ion-to-polymer concentration), so the occupancy fraction in the simulator provides a direct way to translate experimental loading levels into the number of mobile ions represented in the lattice. This mapping is approximate, since the lattice is an abstracted representation rather than a one-to-one correspondence with specific polymer repeat units. So, this total number of cations is then evenly distributed across the vertical rails at the start of the simulation, producing a homogeneous initial ion distribution. From this uniform starting point, the ions are free to migrate in response to electric fields, local concentration gradients, and Coulombic repulsion, allowing the simulation to capture both drift- and diffusion-dominated regimes.

For typical device dimensions (thickness $\approx200nm$, as measured by profilometry), the lattice consists of hundreds of cells along the vertical (thickness) direction and thousands along the lateral dimensions, ensuring sufficient resolution to simulate both short-range hopping events and long-range drift/diffusion processes. Lattice points are spaced such that nearest-neighbor interactions effectively approximate the microstructure of the blended polymer and ion-conducting matrix.

\subsection{Physical Dimensions and Electrode Configuration}

The active layer has a thickness of approximately 209~nm, as determined by profilometry, and an effective active area of 0.0825~cm$^2$, defined by the overlap between the bottom and top electrodes. The simulated geometry reproduces this vertical configuration, with the lower conductive layer serving as the grounded contact and the upper metallic layer acting as the biased electrode, following the structure described by Prado-Socorro \textit{et al.}~\cite{Prado2022}.


In the simulation, the lateral dimensions are scaled to maintain computational efficiency while preserving enough spatial resolution to resolve heterogeneity in cations motion. This balance allows the model to capture meaningful transport phenomena without incurring the computational expense of atomistic simulations.

\subsection{Boundary Conditions Applied at Electrodes and Edges}

At the electrode interfaces, fixed potential (Dirichlet) boundary conditions are applied, with the voltage difference set according to the simulated bias. Cations are free to accumulate near these boundaries, forming double layers and modifying the local electric field, as observed experimentally. At the lateral edges, periodic boundary conditions are used to emulate an extended device area and minimize finite-size artifacts.

\subsection{Table of Physical Constants and Simulation Parameters}

For completeness, all simulation parameters, device settings, and numerical constants are summarized in Supplementary Table S1. The table includes general configurations (such as temperature, computing device, and initial spatial ion distribution), device geometry (grid dimensions), and protocol-specific parameters for the different simulation types (decay of current, hysteresis, and learning/forgetting rate). Each entry provides the parameter name, value, units, and a brief description to facilitate reproducibility of the simulations. By simply changing the \texttt{simulation\_type} parameter, different types of simulations can be performed without modifying other settings. Further information regarding this table and the role of each parameter for the different simulation protocols is provided in subsection S2.2, where we describe the initialization routines used for each simulation type.

\begin{table}[h!]
\centering
\scriptsize
\renewcommand{\arraystretch}{1.2}
\begin{tabular}{l l l p{12cm}}
\hline
\textbf{Variable} & \textbf{Value} & \textbf{Units} & \textbf{Description} \\
\hline
\multicolumn{4}{l}{\textbf{General configurations}} \\
ion\_fraction & 30 & \% & Percentage of ions in the system \\
Temperature & 300 & K & Temperature of the simulation environment \\
save & 1 & - & Flag to save simulation output (1 = yes, 0 = no) \\
dimension\_y & 4000 & - & Simulation grid size in y-direction (number of rails) \\
dimension\_x & 200 & - & Simulation grid size in x-direction (number of cups)\\
starting\_mode & 100 & \% & Percentage of sites initially occupied by ions in the spatial grid\\
simulation\_type & 1 & - & Type of simulation (1=Relaxation decay, 2=Hysteresis, 3=Learning/forgetting rate) \\
relaxation\_time & 0.016 & s & Characteristic relaxation time \\
effective\_voltage\_difference\_factor & 0.00005 & V & Multiplier accounting for voltage effects due to ion–ion repulsion \\
device & cpu & - & Compute device (cpu or gpu) \\

\multicolumn{4}{l}{\textbf{Relaxation decay parameters}} \\
constant\_voltage & 0 & V & Constant voltage applied during relaxation decay \\
polarization\_time & 5 & s & Duration of polarization step \\
polarization\_voltage\_applied & 5 & V & Voltage applied during polarization \\
total\_time & 300 & s & Total simulation time for relaxation decay \\

\multicolumn{4}{l}{\textbf{Hysteresis parameters}} \\
maximum\_voltage\_H & 1 & V & Maximum voltage applied during hysteresis sweep \\
minimum\_voltage\_H & 0 & V & Minimum voltage applied during hysteresis sweep \\
sweep\_rate & 0.25 & V/s & Rate of voltage sweep during hysteresis \\

\multicolumn{4}{l}{\textbf{Learning/forgetting rate parameters}} \\
maximum\_voltage & 1 & V & Maximum voltage applied  \\
minimum\_voltage & -1 & V & Minimum voltage applied \\
time\_maximum\_pulses & 5 & s & Duration of maximum pulses \\
time\_minimum\_pulses & 5 & s & Duration of minimum pulses \\
baseline\_pulse & 0 & V & Baseline voltage of pulses \\
pulses\_shape & 1 & - & Shape type of applied pulses (1 = triangles, 2 = blocks)\\
pulse\_frequency & 10 & Hz & Frequency of the applied pulses \\
\hline
\end{tabular}
\caption{Simulation configuration parameters with values, units, and descriptions.}
\label{tab:yaml_config}
\end{table}

\clearpage

\section{Algorithmic Implementation}

The core simulation framework builds upon the general stochastic update strategy introduced in the work by Gutierrez-Finol, et al. \cite{Gutierrez2023}, where discrete states are updated probabilistically according to local transition rules derived from physical parameters. In our case, however, the underlying physics, spatial representation, and evaluation of the state matrix are fundamentally different. Whereas the lanthanide nanomagnet study models spin states in molecular-scale systems with magnetic interactions, our implementation focuses on the drift–diffusion behavior of ions within a polymer-based memristive matrix under applied electric bias. The ``states matrix'' in our model directly encodes the instantaneous positions of each ion in a discrete lattice that represents the nanoscopic structure of the device. At each Monte Carlo time step, this matrix is updated according to field-dependent transition probabilities, thermally activated reverse hops, and hard-core exclusion rules that prevent site over-occupation. This adaptation retains the computational efficiency and scalability of the original probabilistic bit algorithm, while embedding a transport model tailored to ionic conduction, charge accumulation, and relaxation processes characteristic of soft-matter memristors.

\subsection{Step-by-Step Explanation of the Markov Chain Monte Carlo Framework}

This section describes the stochastic ion-transport model used in our polymer-based memristor simulations. The framework is based on a Markov Chain Monte Carlo (MCMC) scheme, where mobile cations occupy discrete sites in a two-dimensional matrix. Each row represents a one-dimensional rail that constrains ion motion laterally, while multiple rails are stacked vertically to reproduce the layered device geometry. Within a given rail, ions migrate stochastically along the horizontal direction: at each time step, an ion can hop left, hop right, or remain in place, with probabilities determined by the local electric field, temperature, and spatial zone. Inter-rail hopping and vertical motion are not included in the present implementation. The main stages of the simulation loop are as follows:

\subsubsection{Initialization}

The device geometry is mapped onto a discrete lattice where each occupied site corresponds to the position of a single ion. 
\begin{itemize}
    \item[-] The vertical dimension ($y$) corresponds to the number of rails (rows).
    \item[-] The horizontal dimension ($x$) corresponds to the number of available hopping sites per rail.
    \item[-] The total number of ions is determined from the experimental ion fraction data and uniformly distributed at the start.
\end{itemize}

\subsubsection{Electric-field-dependent energy gap}

At each time step, the influence of the applied bias is translated into an \textit{energy gap} per possible hop:
\begin{equation} \label{eq:Delta_E}
\Delta E = \frac{q_{\mathrm{ion}} \cdot \Delta V_{\mathrm{net}} \cdot d_{hop}}{t_{dev}}
\end{equation}
where:
\begin{itemize}
    \item[-] $q_{\mathrm{ion}}$ is the ionic charge of the ion (e),
    \item[-] $\Delta V_{\mathrm{net}}$ is the net potential drop experienced across the device in volts (including the effect of prior ion accumulation),
    \item[-] $d_{hop}$ is the hop distance between lattice sites (nm),
    \item[-] $t_{dev}$ is the device thickness (nm).
\end{itemize}

This converts the electric field effect into an energy gap ($\Delta E$ in eV) between forward and backward moves determining their asymmetry.

\subsubsection{Boltzmann Factor for Thermal Activation}

Thermal activation and field effects are incorporated using a Boltzmann factor:
\begin{equation} \label{eq:Boltzmann}
B_{zone} = \exp\left( - \frac{\Delta E_{zone}}{k_{B} T} \right)
\end{equation}
where $k_{B}$ is the Boltzmann constant ($8.6173\times10^{-5}$ eV/K), $T$ is the absolute temperature, and the subscript $zone$ denotes the spatial ``zone'' index (1 to 10) used to account for lateral inhomogeneities. Each zone has its own $B_{zone}$ value, meaning regions of the device with different field strengths or ion densities will have different directional bias.

\subsubsection{Directional hopping probabilities}

The Boltzmann weighting is combined with a baseline hopping probability $P_{\mathrm{0}}$ (set by material-specific diffusion constants and time-step scaling) to produce left and right hopping probabilities:
\begin{equation} \label{eq:P_left}
P^{(\leftarrow)}_{zone} = \frac{P_{\mathrm{0}}}{1 + B_{zone}}
\end{equation}

\begin{equation} \label{eq:P_right}
P^{(\rightarrow)}_{zone} = B_{zone} \cdot P^{(\leftarrow)}_{zone}
\end{equation}

These expressions ensure that forward bias increases the probability of hopping in the field direction, while reverse hops remain possible via thermal activation.

\subsubsection{Spatial Zones and Local Probabilities}

The simulated polymer-based memristive device is represented by a discrete lattice of size 
$N_y \times N_r$, where $N_y$ is the number of horizontal rows (``rails'') and $N_r$ is the number of ions per row. 
The position of each ion is stored in the integer-valued matrix:
\[
\mathbf{X}(t) \in \mathbb{Z}^{N_y \times N_r}
\]
where each element $X_{ij}(t)$ denotes the column index (spatial position) of ion $j$ in row $i$ at time step $t$.

As it was mentioned before, the lattice is divided into $Z$ vertical zones of equal width, such that the zone index for ion $(i,j)$ is:
\[
z_{ij} = \left\lceil \frac{X_{ij}(t)}{L_x / Z} \right\rceil
\]
where $L_x$ is the total number of discrete columns in the horizontal direction.  
Each zone $z$ is associated with a pair of directional hopping probabilities (equations \ref{eq:P_left} and \ref{eq:P_right}), which are precomputed from the electric-field-dependent Boltzmann statistics.

\subsubsection{Random Sampling and Exclusion Rules}

At each time step, two independent random matrices are generated:
\[
R^{(\rightarrow)}_{ij}, \quad R^{(\leftarrow)}_{ij} \sim \mathcal{U}(0,1)
\]
for all $i \in \{1,\dots,N_y\}$ and $j \in \{1,\dots,N_r\}$.

The minimum distances to the nearest occupied neighbor are computed as:
\[
\Delta^{(\rightarrow)}_{ij} = X_{i,j+1}(t) - X_{ij}(t), \quad
\Delta^{(\leftarrow)}_{ij} = X_{ij}(t) - X_{i,j-1}(t)
\]
with appropriate handling of the first and last ions in each row to account for boundary conditions.  

A movement is only considered valid if:
\[
\Delta^{(\rightarrow)}_{ij} > 1 \quad \text{(for right moves)}, \quad
\Delta^{(\leftarrow)}_{ij} > 1 \quad \text{(for left moves)}
\]

\subsubsection{Movement Acceptance and Position Update}

Define the movement masks:
\[
M^{(\rightarrow)}_{ij} = 
\begin{cases}
1, & \text{if } R^{(\rightarrow)}_{ij} \leq p^{(\rightarrow)}_{z_{ij}} \ \text{and} \ \Delta^{(\rightarrow)}_{ij} > 1 \\
0, & \text{otherwise}
\end{cases}
\]
\[
M^{(\leftarrow)}_{ij} = 
\begin{cases}
1, & \text{if } R^{(\leftarrow)}_{ij} \leq p^{(\leftarrow)}_{z_{ij}} \ \text{and} \ \Delta^{(\leftarrow)}_{ij} > 1 \\
0, & \text{otherwise}
\end{cases}
\]

The ion positions are updated according to:
\[
X_{ij}(t+1) = X_{ij}(t) + M^{(\rightarrow)}_{ij} - M^{(\leftarrow)}_{ij}
\]
This procedure is applied simultaneously to all ions, ensuring a parallel Monte Carlo update at each iteration. In summary, for each ion:

\begin{enumerate}
    \item A random number $r^{(\rightarrow)}$ is generated for the right hop trial; if $r^{(\rightarrow)} \leq P^{(\rightarrow)}_{zone}$ and the neighboring site is empty, the ion moves right.
    \item Similarly, a random number $r^{(\leftarrow)}$ is compared to $P^{(\leftarrow)}_{zone}$ for left movement.
    \item Movement is only allowed if the target position is unoccupied and within the device boundaries.
\end{enumerate}
The movement conditions also include spatial constraints from the \textit{current ion occupancy}. This means that high-density regions naturally reduce mobility through site-blocking, a discrete analog to space-charge limitation.

\subsubsection{Implicit space-charge effects}

Although no explicit Poisson equation is solved, the simulation incorporates space-charge feedback because:
\begin{itemize}
    \item[-] The net voltage difference $\Delta V_{\mathrm{net}}$ used in Eq.~(1) depends on how ions redistribute during the simulation.
    \item[-] Ion accumulation near electrodes reduces the local field in subsequent steps, lowering the drift probability in those regions.
    \item[-] This creates a dynamic coupling between ionic configuration and movement statistics, mimicking the screening and repulsion observed experimentally.
\end{itemize}

\subsubsection{Summary of drift, diffusion, and space-charge representation}
\begin{itemize}
    \item[-] Drift: Introduced through the energy gap term proportional to the applied bias.
    \item[-] Diffusion: Captured via the Boltzmann factor allowing hopping against the field.
    \item[-] Space charge: Implicitly handled by occupancy-dependent field updates and site-blocking rules.
\end{itemize}

This combination provides a computationally efficient yet physically grounded way to reproduce experimentally observed ionic behaviors such as relaxation, hysteresis, and retention decay, without resorting to full atomistic simulations.

\subsection{Description of initialization routines for each protocol simulation}

Before running any protocol, the user must specify a set of initialization parameters in the configuration file (\texttt{user\_configurations.yaml}). These parameters determine the experimental conditions, simulation geometry, and time--voltage characteristics for the selected protocol. The initialization stage ensures that all simulations are reproducible and based on clearly defined input conditions.

\subsubsection{General Configuration Parameters}

These parameters are required for all protocol types:
\begin{itemize}
    \item[-] \textbf{Ion Occupancy Fraction (0--100\%)} -- Defines the proportion of available lattice sites initially occupied by mobile ions. This parameter is derived from the experimental ion concentration, typically expressed as the ratio of ionic species to the ion-conducting polymer matrix. It provides a direct mapping from experimental loading levels to the number of cations represented in the simulated device.
    \item[-] \textbf{Temperature (K)} -- Operating temperature in Kelvin. Although the current implementation uses a user-defined relaxation time rather than temperature-dependent mechanisms, this value is preserved for compatibility and potential thermally dependent simulations.
    \item[-] \textbf{Save Results (1 or 0)} -- Enables or disables the saving of simulation results to file.
    \item[-] \textbf{Device Dimensions:}
    \begin{itemize}
        \item \emph{Dimension in y} -- Number of parallel rails (rows) in the device.
        \item \emph{Dimension in x} -- Number of available ion sites (columns) along each rail.
    \end{itemize}
    \item[-] \textbf{Starting Mode (0--100\%)} -- Specifies the uniformity of the initial ion distribution. A value of 100\% indicates fully uniform occupancy, whereas lower values produce more spatial variation.
    \item[-] \textbf{Type of Simulation} -- Selection of the simulation protocol:
    \begin{enumerate}
        \item Current decay due to relaxation of electric polarization
        \item Hysteresis
        \item Learning/forgetting rate
    \end{enumerate}
    \item[-] \textbf{Relaxation Time (s)} -- The intrinsic timescale over which ionic rearrangements occur.
    \item[-] \textbf{Effective Voltage Difference Factor} -- Scaling factor applied to the applied voltage to account for charge accumulation effects.
\end{itemize}

\subsubsection{Decay of current due to relaxation of electric polarization}

When \emph{Current decay} is selected, the following parameters must be specified:
\begin{itemize}
    \item[-] \textbf{Constant Voltage (V)} -- Voltage applied during the relaxation phase.
    \item[-] \textbf{Polarization Time (s)} -- Duration of the pre-polarization stage before the relaxation begins.
    \item[-] \textbf{Polarization Voltage (V)} -- Voltage used during the polarization stage.
    \item[-] \textbf{Total Experiment Time (s)} -- Duration of the entire relaxation experiment.
    \item[-] \textbf{Maximum and Minimum Voltage (V)} -- Define the operating range; in constant-voltage decay, these values may be used for later analysis.
\end{itemize}

\subsubsection{Hysteresis (Variable Voltage)}

For the \emph{Hysteresis} protocol, the user specifies:
\begin{itemize}
    \item[-] \textbf{Maximum Voltage (V)} -- Peak voltage during the sweep.
    \item[-] \textbf{Minimum Voltage (V)} -- Lowest voltage during the sweep.
    \item[-] \textbf{Sweep Rate (V/s)} -- Voltage change rate that defines the period of the sweep cycle. This parameter determines the total duration of the experiment according to the voltage range.
\end{itemize}

\subsubsection{Learning / Forgetting Rate}

In this protocol, the applied voltage consists of pulses with varying amplitude and polarity. Required parameters are:
\begin{itemize}
    \item[-] \textbf{Maximum Voltage (V)} -- Pulse amplitude during the ``learning'' phase.
    \item[-] \textbf{Minimum Voltage (V)} -- Pulse amplitude during the ``forgetting'' phase.
    \item[-] \textbf{Time for Maximum Pulses (s)} -- Duration of the learning phase.
    \item[-] \textbf{Time for Minimum Pulses (s)} -- Duration of the forgetting phase.
    \item[-] \textbf{Baseline Pulse} -- Offset voltage applied between pulses.
    \item[-] \textbf{Pulse Shape} -- Either \emph{Triangular} or \emph{Rectangular}, defining the waveform within each pulse cycle.
    \item[-] \textbf{Pulse Frequency (Hz)} -- Number of pulses per second.
\end{itemize}

\subsubsection{Role of Initialization}

The initialization routine reads these values in the order they appear in the configuration file and constructs the internal \texttt{config} structure. This structure defines:
\begin{itemize}
    \item[-] The spatial arrangement and number of ions (\texttt{dimension\_x}, \texttt{dimension\_y}, \texttt{ion\_fraction}).
    \item[-] The timescale and number of simulation steps (\texttt{relaxation\_time}, \texttt{total\_time}).
    \item[-] The complete voltage vs.\ time profile for the selected protocol.
\end{itemize}

By providing a complete and well-defined set of parameters, the user ensures that the simulation can be exactly reproduced and that comparisons between different experiments are valid.

\section{Conversion of Ionic Distributions to Conductance/Current Variables}

To enable direct comparison between simulated ion dynamics and experimental observables, we developed a function that converts the ionic occupation data into quantities proportional to conductance and current. The procedure is based on tracking the fraction of ions located in specific deciles of the simulated lattice, which are assumed to dominate charge transport pathways during device operation. This choice reflects the experimental observation that conductance in polymer-based memristive devices is strongly correlated with ionic accumulation near interfaces and preferential migration channels.

For hysteresis simulations, the algorithm identifies the fraction of ions located in the first and tenth deciles of the lattice, depending on the sign of the applied bias. When the voltage is positive, the lowest decile (near one electrode) is used, while under negative voltage the uppermost decile is considered. This decile-dependent ionic ratio is normalized by the total ion number and multiplied by the applied time vector, yielding a trajectory that can be directly related to device conductance and current–voltage characteristics.

For relaxation-decay simulations, the function extracts the ionic ratio from the first decile as a function of time, which represents the progressive redistribution of ions after a bias pulse. The simulated data are normalized to match the range of experimental delay-time curves obtained from independent measurements, enabling a consistent comparison of decay dynamics between simulation and experiment.

Finally, in general simulation modes, the function records the ionic ratio across the selected deciles together with the applied voltage, producing datasets that can be exported for post-processing or fitting to experimental traces. In all cases, the resulting output provides three data streams: (i) the experimental ion ratio for direct reference, (ii) the original simulated ionic ratio, and (iii) a normalized version rescaled to the experimental range. This workflow allows systematic conversion of raw ionic density profiles into transport-related observables, bridging the gap between microscopic stochastic simulations and macroscopic electrical measurements.

\section{CPU–GPU performance and energy efficiency comparison}

To quantitatively assess the benefits of GPU acceleration, we benchmarked the JAX implementation of the simulation on both CPU and GPU backends using identical workloads. Tables~\ref{tab:cpu_gpu_performance} and ~\ref{tab:cpu_gpu_energy} summarize the runtime, computational throughput, and energy metrics for increasing numbers of simulated rails.

Overall, the GPU consistently achieved higher throughput and lower energy consumption across all test cases. The performance speedup ranged from approximately 1.3× for small-scale simulations (500 rails) to about 17× for large-scale workloads (128{,}000 rails). Correspondingly, the energy efficiency improved significantly, with up to sevenfold energy savings at higher scales. These improvements arise from the higher degree of parallelization and better computational intensity per watt of modern GPUs, particularly when combined with JAX’s just-in-time (JIT) compilation and vectorized operation support.

\begin{table}[h!]
\centering
\scriptsize
\renewcommand{\arraystretch}{1.2}
\resizebox{\textwidth}{!}{
\begin{tabular}{rccccccc}
\hline
\textbf{Rails} & \textbf{CPU time} & \textbf{GPU time} & \textbf{CPU steps} & \textbf{GPU steps} & \textbf{CPU rails} & \textbf{GPU rails} & \textbf{Speedup} \\
 & (s) & (s) & (steps/s) & (steps/s) & (rails/s) & (rails/s) &  \\
\hline
500 & 125 & 94 & 1498 & 1991 & 748800 & 995745 & 1.3 \\
1000 & 210 & 100 & 891 & 1872 & 891429 & 1872000 & 2.1 \\
2000 & 401 & 122 & 467 & 1534 & 933666 & 3068852 & 3.3 \\
4000 & 866 & 162 & 216 & 1156 & 864665 & 4622222 & 5.3 \\
8000 & 1987 & 224 & 94 & 836 & 753699 & 6685714 & 8.9 \\
16000 & 4130 & 359 & 45 & 521 & 725230 & 8343175 & 11.5 \\
32000 & 9277 & 635 & 20 & 295 & 645726 & 9433701 & 14.6 \\
64000 & 18932 & 1165 & 10 & 161 & 632833 & 10283948 & 16.3 \\
128000 & 37288 & 2224 & 5 & 84 & 642609 & 10774101 & 16.8 \\
\hline
\end{tabular}
}
\caption{Performance comparison between CPU and GPU implementations for increasing numbers of rails. The GPU consistently achieves higher throughput and speedup as the workload scales.}
\label{tab:cpu_gpu_performance}
\end{table}

\begin{table}[h!]
\centering
\scriptsize
\renewcommand{\arraystretch}{1.2}
\resizebox{\textwidth}{!}{
\begin{tabular}{rcccccccc}
\hline
\textbf{Rails} & \textbf{CPU power} & \textbf{GPU power} & \textbf{CPU energy} & \textbf{GPU energy} & \textbf{CPU energy/rail} & \textbf{GPU energy/rail} & \textbf{Energy savings} \\
 & (W) & (W) & (Wh) & (Wh) & (mWh) & (mWh) &  \\
\hline
500 & 20.24 & 31.4 & 0.70 & 0.82 & 1.41 & 1.64 & 0.9 \\
1000 & 19.57 & 32.8 & 1.14 & 0.91 & 1.14 & 0.91 & 1.3 \\
2000 & 18.87 & 32.53 & 2.10 & 1.10 & 1.05 & 0.55 & 1.9 \\
4000 & 18.46 & 32.75 & 4.44 & 1.47 & 1.11 & 0.37 & 3.0 \\
8000 & 17.32 & 34.8 & 9.56 & 2.17 & 1.19 & 0.27 & 4.4 \\
16000 & 17.29 & 36.89 & 19.84 & 3.68 & 1.24 & 0.23 & 5.4 \\
32000 & 18.76 & 38.81 & 48.34 & 6.85 & 1.51 & 0.21 & 7.1 \\
64000 & 18.03 & 43.03 & 94.82 & 13.92 & 1.48 & 0.22 & 6.8 \\
128000 & 17.95 & 43.92 & 185.92 & 27.13 & 1.45 & 0.21 & 6.9 \\
\hline
\end{tabular}
}
\caption{Energy and power comparison between CPU and GPU implementations. GPU acceleration significantly reduces energy per rail and overall energy consumption as problem size increases.}
\label{tab:cpu_gpu_energy}
\end{table}

\section{Model Validation and Parameter Dependence}

\subsection{Effect of Rail Number on Signal Noise}

\begin{figure}[h!]
    \centering
    \includegraphics[width=1.0\linewidth]{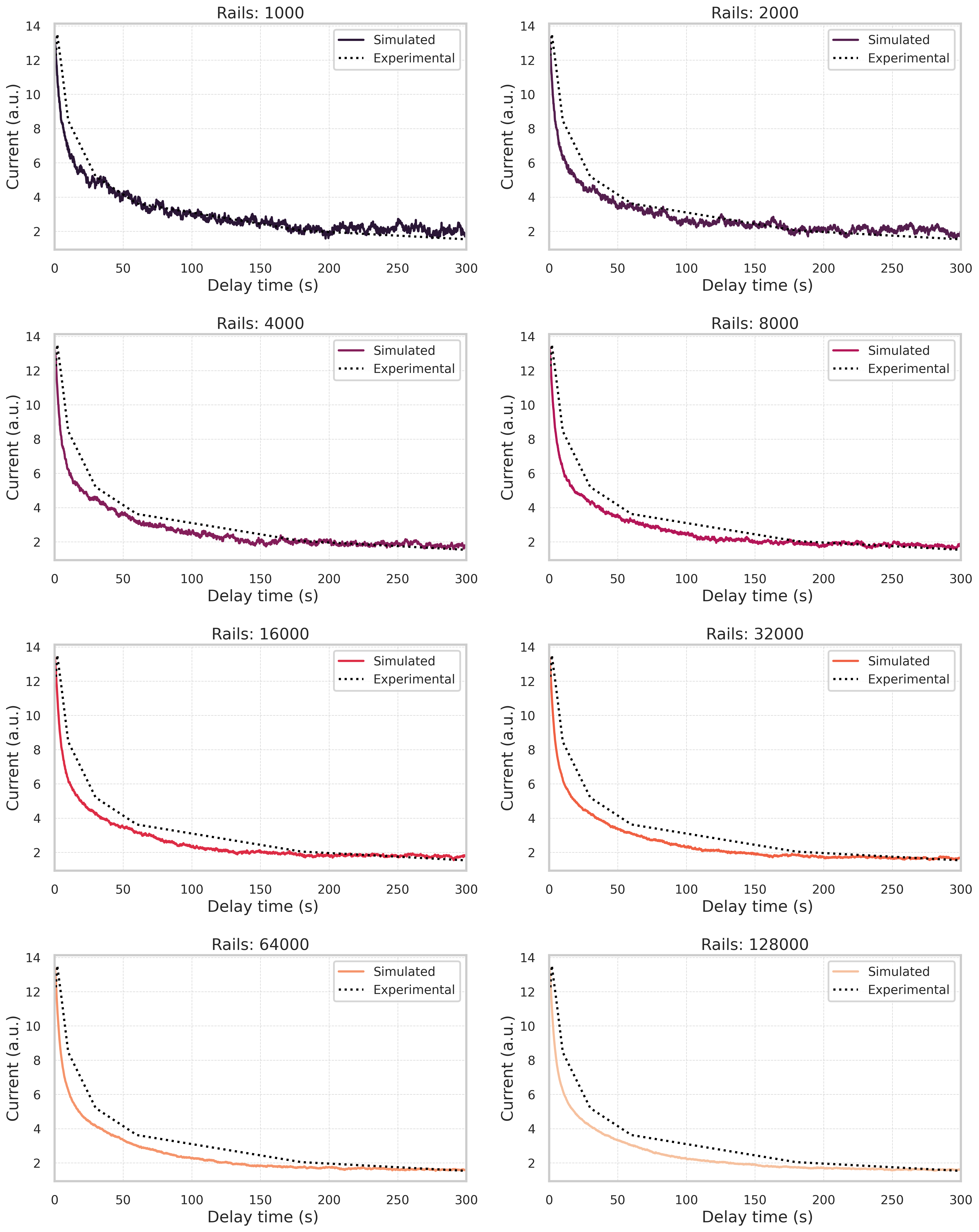}
    \caption{Figure~\ref{fig:hysteresis_sweep_rates_SI} shows the simulated I--V hysteresis loops. At increasing voltage sweep rates, the simulated I--V curves exhibit a progressive reduction in current and loop amplitude. This trend arises from the limited response time of the system: faster voltage changes restrict the extent of ionic and electronic rearrangements, thereby diminishing the overall current magnitude and narrowing the hysteresis loops. In contrast to the simulations presented in the main text, where each sweep begins from the final ionic configuration of the preceding one to capture cumulative memory effects, the hysteresis loops shown here are fully independent, with each simulation starting from an random initial configuration. This separation isolates the intrinsic effect of the sweep rate from any history-dependent contributions.
}
    \label{fig:delaytime_rails}
\end{figure}

Increasing the number of rails in the simulation effectively reduces the visual noise in the current curves (Figure~\ref{fig:delaytime_rails}). Each rail contributes independently to the overall signal, so random fluctuations present in individual rails are averaged out. As a result, the curves become smoother, more reproducible, and the underlying relaxation behavior is more clearly observed.

\subsection{Current relaxation and influence of relaxation time}

When the polarization voltage is removed, the driving electric field that sustains ionic displacement across the nanodevice vanishes. As a result, mobile ions gradually relax back toward their equilibrium distribution within the material. This ionic redistribution reduces the internal electric field and charge imbalance responsible for the measured current. Consequently, the current decays over time until reaching a steady-state value determined by the equilibrium ionic configuration, reflecting the intrinsic relaxation dynamics of the device.

\begin{figure}[h!]
    \centering
    \includegraphics[width=1.0\linewidth]{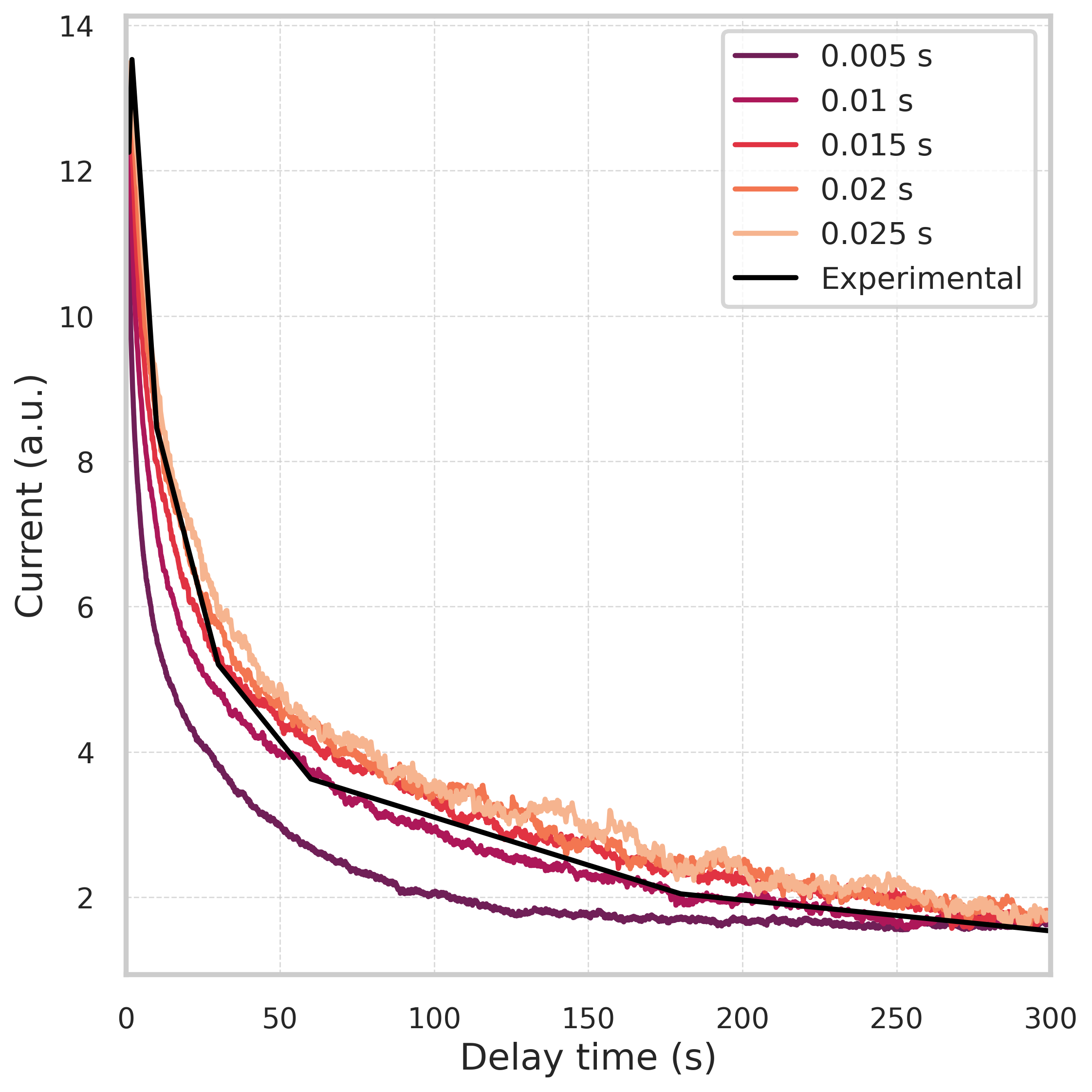}
    \caption{Simulated relaxation decay curves obtained at different relaxation times ($0.005$--$0.025$~s), compared with experimental data (solid black line). Each simulated curve corresponds to a distinct relaxation parameter in the model. Shorter relaxation times produce faster decay dynamics and lower steady-state currents, while longer times lead to a slower relaxation toward equilibrium. The experimental trace serves as a reference for evaluating the temporal response of the simulated system.}
    \label{fig:relaxation_decay_multiple_times}
\end{figure}

Additionally, as shown in Figure~\ref{fig:relaxation_decay_multiple_times}, the visual noise in the simulated current decreases with increasing relaxation time. Longer relaxation times allow the ionic and electronic variables to evolve more smoothly, effectively filtering out high-frequency fluctuations and producing cleaner, more stable decay profiles. This behavior highlights the damping effect of slower ion dynamics on transient current noise within the nanodevice.

\subsection{Hysteresis loops at different voltage sweep rates}

\begin{figure}[h!]
    \centering
    \includegraphics[width=1.0\linewidth]{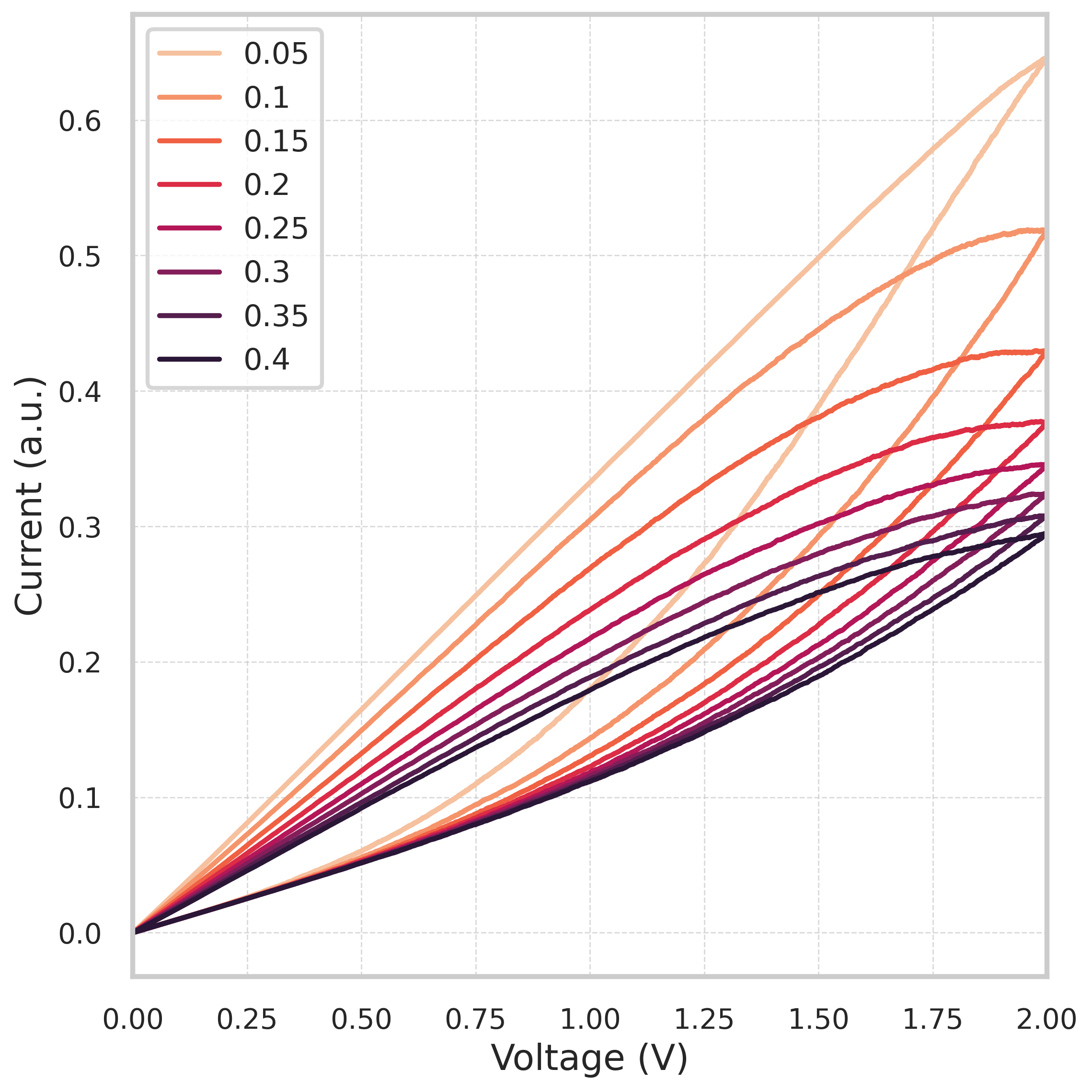}
    \caption{Simulated I--V hysteresis loops obtained at different voltage sweep rates (0.05--0.4 V/s). Each curve corresponds to a specific sweep rate, indicated in the legend. As the sweep rate increases, the overall current decreases, resulting in smaller loop amplitudes. This behavior reflects the dynamic response of the system, where faster sweeps reduce the time available for ionic or electronic rearrangements.}
    \label{fig:hysteresis_sweep_rates_SI}
\end{figure}

Figure~\ref{fig:hysteresis_sweep_rates_SI} shows the simulated I--V hysteresis loops. At increasing voltage sweep rates, the simulated I--V hysteresis loops exhibit a progressive reduction in current and loop amplitude. This trend arises from the limited response time of the system: faster voltage changes restrict the extent of ionic and electronic rearrangements, thereby diminishing the overall current magnitude and narrowing the hysteresis loops.

\subsection{Relaxation Dynamics of Conductance}

\begin{figure}[h!]
    \centering
    \includegraphics[width=0.8\textwidth]{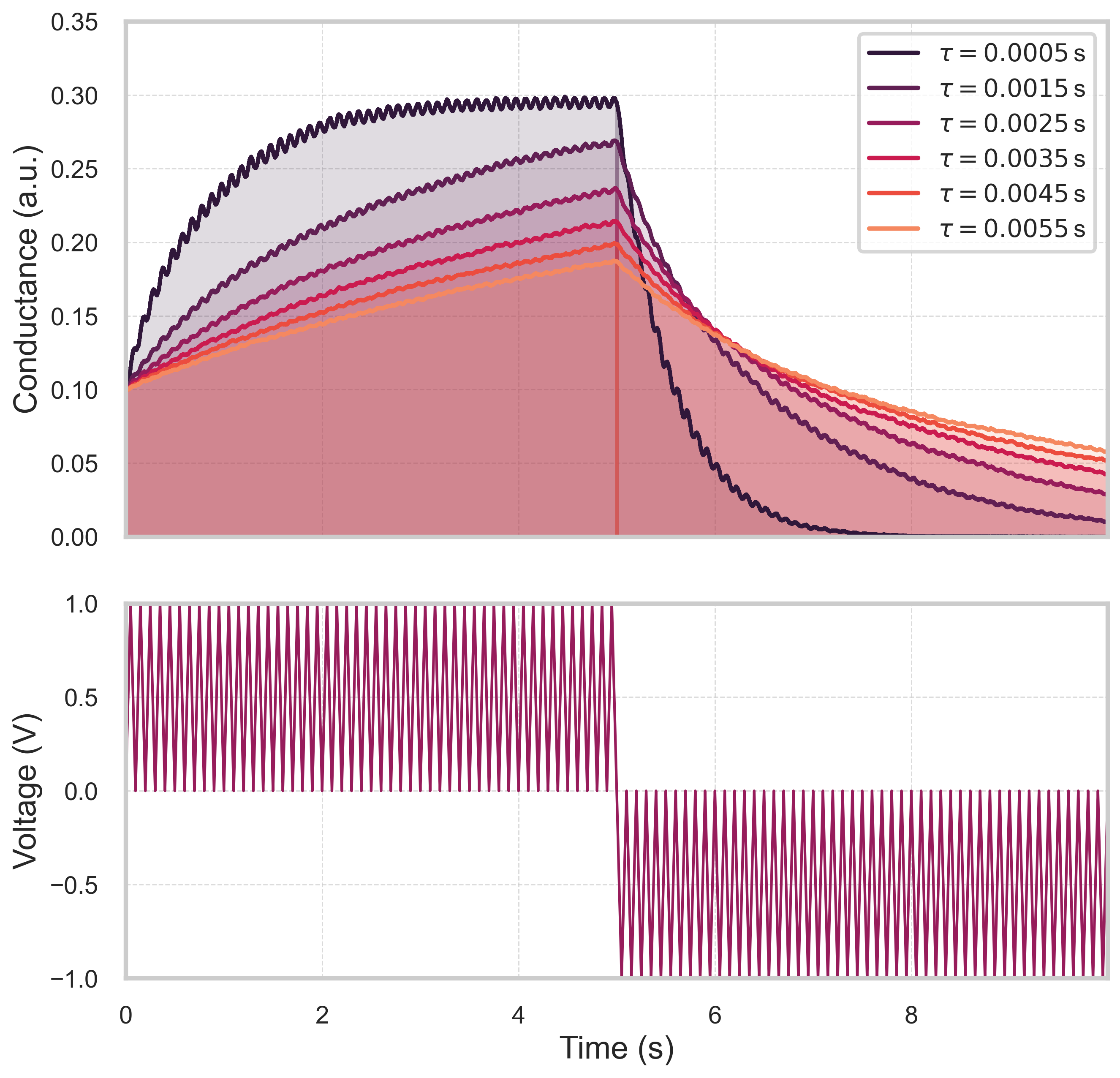}
    \caption{Time evolution of conductance (top) for six different relaxation times, $\tau$, and corresponding applied voltage waveform (bottom). Shaded areas highlight the integrated conductance during the learning and forgetting phases.}
    \label{fig:L_F_relaxation_combined_conductance_5curves}
\end{figure}

Figure~\ref{fig:L_F_relaxation_combined_conductance_5curves} shows the time evolution of the conductance for the device under a pulsed voltage waveform, highlighting the effect of different relaxation times, $\tau$. The top panel displays the conductance traces for five different $\tau$ values ranging from $0.0005$~s to $0.0055$~s. Each trace exhibits a characteristic learning (increase) phase during the initial positive voltage pulse, followed by a forgetting (decay) phase once the voltage reverses. 

Shaded areas represent the regions under each conductance curve, with transparency adjusted to allow visualization of overlapping curves. As expected, shorter relaxation times lead to a faster rise and decay of the conductance, while longer relaxation times result in more gradual changes, consistent with a slower adaptation to the applied voltage.

The bottom panel presents the corresponding voltage waveform applied to the device, which consists of a high-frequency square-like signal alternating between positive and negative values. The alignment of the conductance response with the voltage waveform demonstrates the device's temporal dynamics and memory effects under rapid voltage modulation.

\bibliographystyle{vancouver}
\bibliography{bibliography}